\documentclass{amsart}
\newtheorem{thm}{Theorem}
\newtheorem{prop}{Proposition}
\newtheorem{cor}{Corollary}
\newtheorem{Def}{Definition}
\newtheorem{lem}{Lemma}
\usepackage{amsmath,amssymb}
\usepackage{enumerate}
\usepackage{latexsym,amscd,amsfonts}
\def \Z {\mathbb Z}
\def \R {\mathbb R}
\def \C {\mathbb C}

\def \N {\mathbb N}

\def \P {\mathbb P}
\def \E {\mathbb E}
\providecommand{\dd}{\,\mathrm{d}}
\providecommand{\ee}{\mathrm{e}}

\DeclareMathOperator{\Sp}{Sp}
\DeclareMathOperator{\supp}{supp}
\begin{document}
\title{H\"older continuity of the IDS for matrix-valued Anderson models}
\author{Hakim \textsc{Boumaza} }
\address{Keio University, Department of Mathematics\\
Hiyoshi 3-14-1\\
Kohoku-ku 223-8522\\
Yokohama\\
Japan\\}
\email{boumaza@math.keio.ac.jp}
\thanks{The author is supported by JSPS Grant P07728}
\date{\today}

\begin{abstract}
We study a class of continuous matrix-valued Anderson models acting on $L^{2}(\R^{d})\otimes \C^{N}$. We prove the existence of their Integrated Density of States for any $d\geq 1$ and $N\geq 1$. Then for $d=1$ and for arbitrary $N$, we prove the H\"older continuity of the Integrated Density of States under some assumption on the group $G_{\mu_{E}}$ generated by the transfer matrices associated to our models. This regularity result is based upon the analoguous regularity of the Lyapounov exponents associated to our model, and a new Thouless formula which relates the sum of the positive Lyapounov exponents to the Integrated Density of States. In the final section, we present an example of matrix-valued Anderson model for which we have already proved, in a previous article, that the assumption on the group $G_{\mu_{E}}$ is verified. Therefore the general results developed here can be applied to this model.
\end{abstract}


\maketitle

\section{Introduction}

We will study the question of the existence of the Integrated Density of States and its regularity for continuous matrix-valued Anderson models of the form :
\begin{equation}\label{modeld}
H_{A}(\omega)=-\Delta_{d}\otimes I_{N}+\sum_{n\in \Z^{d}} V_{\omega}^{(n)}(x-n)
\end{equation}
acting on $L^{2}(\R^{d})\otimes \C^{N}$, $d$ and $N$ are non-negative integers, $I_{N}$ is the identity matrix of order $N$ and $\Delta_{d}$ denotes the $d$-dimensional continuous Laplacian.
Let $(\Omega, \mathcal{A}, \mathsf{P})$ be a complete probability space and $\omega \in \Omega$. For every $n\in \Z$, the functions $x\mapsto V_{\omega}^{(n)}(x)$ will be symmetric matrix-valued functions, supported in $[0,1]^{d}$, and bounded uniformly on $x,n$ and $\omega$. We also set : 
$$\forall x\in \R^{d},\ V_{\omega}(x)=\sum_{n\in \Z^{d}} V_{\omega}^{(n)}(x-n)$$ and denote by $V_{\omega}$ the maximal multiplication operator by $x\mapsto V_{\omega}(x)$. The function 
$x\mapsto V_{\omega}(x)$ is uniformly bounded on $\R$ in $x$ and in $\omega$. The potential $V_{\omega}$ will also be such that the operator $H_{A}(\omega)$ is $\Z^{d}$-ergodic. As a bounded perturbation of $-\Delta_{d}\otimes I_{N}$, the operator $H_{A}(\omega)$ is self-adjoint on the Sobolev space $H^{2}(\R^{d})\otimes \C^{N}$. 
\vskip 3mm

We want to define a function of the real variable  which will count the number of proper energy states of $H_{A}(\omega)$ below a fixed energy $E$. For systems like (\ref{modeld}), such a definition will usually lead to an infinite function as the operators we study act on an infinite-dimensional Hilbert space and thus have infinitely many spectral values. To avoid this problem, we will define our function, the Integrated Density of States or IDS, as a thermodynamical limit as explained in Section \ref{section2}. It will lead to a problem of existence of such a thermodynamical limit. We will prove the existence of the IDS in Section \ref{section2} for any $d$ and any $N$. This existence proof will be based upon a matrix-valued Feynman-Kac formula proven in \cite{boulton} and the adaptation of the argument of Carmona in \cite{carmonastflour} to matrix-valued operators. Once we have proven the existence of the IDS, we will study its regularity as a function of the energy parameter $E$. For this second step, we will restrict ourselves to the case where $d=1$ and $N$ is arbitrary, to be able to use the tools coming from the theory of ODE such as the notion of a transfer matrix. We will prove in Section \ref{section4} that under some assumption on $V_{\omega}$, or more precisely on the group generated by the transfer matrices associated to $H_{A}(\omega)$, the IDS is locally H\"older continuous. This result will come from the analoguous regularity result on Lyapounov exponents proved in Section \ref{section3}, and from a Thouless formula proven in Section \ref{section4} which relates the IDS to the Lyapounov exponents. To prove this Thouless formula, we use results of Kotani and Simon in \cite{kotanis} and Kotani in \cite{kotani83}. The regularity result on Lyapounov exponents is based upon the results of Carmona and Lacroix in \cite{carmona} and Lacroix, Klein and Speis in \cite{KLS}. We also need to prove estimates on the transfer matrices for our model (\ref{modeld}) (for $d=1$) similar to those proven in \cite{DSS02} in the scalar-valued case. In a final section, we present an example of continuous matrix-valued Anderson model for which the needed assumption on the group generated by the transfer matrices is verified. This example is the following matrix-valued Anderson-Bernoulli model :
\begin{equation}\label{model2}
H_{AB}(\omega)=-\frac{\dd^{2}}{\dd x^{2}}\otimes I_{2} +\left( \begin{array}{cc}
0 & 1 \\ 
1 & 0
\end{array}\right)+\sum_{n\in \Z} \left( \begin{array}{cc}
\omega_{1}^{(n)} \chi_{[0,1]}(x-n) & 0 \\
0 & \omega_{2}^{(n)} \chi_{[0,1]}(x-n)
\end{array}\right)
\end{equation}
acting on $L^{2}(\R)\otimes \C^{2}$, with $(\omega_{1}^{(n)})_{n\in \Z}$ and $(\omega_{2}^{(n)})_{n\in \Z}$ two independent sequences of independent and identically distributed (\emph{i.i.d.}) random variables with common law $\nu$ such that $\{ 0,1\} \subset \supp \nu$. This model has already been studied by the author in \cite{boumaza} as an improvement of a result by Stolz and the author in \cite{stolzboumaza}. We proved in \cite{stolzboumaza} absence of absolutely continuous spectrum and pointed out that the improvement made in \cite{boumaza} was necessary to be able to prove local H\"older continuity of the IDS.
\vskip 3mm

The study of the regularity of the IDS is an important step to prove Anderson localization by using a multiscale analysis scheme. It is the key ingredient to prove a Wegner estimate as was done in \cite{CKM} and to adapt it to the case of scalar-valued continuous Anderson model in \cite{DSS02}. We believe that once we will have adapted existing multiscale analysis schemes to the case of matrix-valued operators then it will be possible to prove a Wegner estimate and an Initial Length Scale Estimate for model (\ref{modeld}) for $d=1$ and arbitrary $N$. We will then be able to prove Anderson and dynamical localization for this model as explained in \cite{stollmann}.
\vskip 3mm

The question of localization for one-dimensional continuous matrix-valued Anderson model is coming from a more general problem on Anderson models. Localization for continuous Anderson models in dimension $d\geq 2$ at all energies is still an open problem if one looks for arbitrary disorder, including Bernoulli randomness. A possible approach to the localization for $d=2$ is to discretize one direction, which leads to considering a one-dimensional Anderson model, no longer scalar-valued, but $N\times N$ matrix-valued as here for $d=1$. What is already well understood is the case of dimension one scalar-valued continuous Schr\"odinger operators with arbitrary randomness (see \cite{DSS02}) and discrete matrix-valued Schr\"odinger operators, also for arbitrary randomness (see \cite{GM89} and \cite{KLS}). We want to combine here techniques of \cite{DSS02} and \cite{KLS} to get the local H\"older continuity of the IDS for continuous matrix-valued models.

We finish by mentioning that different methods have been used in \cite{KMPV} to prove localization properties for random operators on discrete strips. They are based upon the use of spectral averaging techniques which did not allow to handle with singular distributions of the random parameters like in our model (\ref{model2}).

\section{Existence of the IDS}\label{section2}

In this section we will define the IDS associated to the operator $H_{A}(\omega)$ and prove its existence. The proof of the existence for the IDS will strongly relie on a matrix-valued Feynman-Kac formula which we will present after the definition of the IDS.

As we have already noticed in the introduction, the operator $H_{A}(\omega)$ is self-adjoint and $\Z^{d}$-ergodic. But, in some parts of the following proofs, and also in Section \ref{section4}, we will need a stronger assumption of $\R^{d}$-ergodicity for $H_{A}(\omega)$ instead of only $\Z^{d}$-ergodicity. To avoid this lack of $\R^{d}$-ergodicity in general, we can refer to the suspension procedure developed by Kirsch in \cite{kirsch}. This procedure allows us to construct from $H_{A}(\omega)$ an operator $\tilde{H}_{A}(\tilde{\omega})$, defined on a bigger probability space, which is $\R^{d}$-ergodic. $\tilde{H}_{A}(\tilde{\omega})$ is also constructed in a way such that its IDS and Lyapounov exponents exist if and only if those of $H_{A}(\omega)$ exist, and in this case they are equal for both operators. Considering the use of this suspension procedure we will work in the following with $H_{A}(\omega)$ as if it is $\R^{d}$-ergodic instead of being only $\Z^{d}$-ergodic.

\subsection{Definition of the IDS}

We aim at defining a function that will gives us the mean number per unit volume of spectral values of $H_{A}(\omega)$ situated below a fixed real number $E$. In order to define this function we will first restrict $H_{A}(\omega)$ to cubes of finite volume of $\R^{d}$. Let $L$ be a strictly positive integer and $D=[-L,L]^{d}\subset \R^{d}$ be the cube centered at $0$ and of length $2L$. We set :
\begin{equation}\label{modeldD}
H_{A}^{(D)}(\omega)=-\Delta_{d}^{(D)}\otimes I_{N}+\sum_{n\in \Z^{d}} V_{\omega}^{(n)}(x-n) 
\end{equation}
the restriction of $H_{A}(\omega)$ acting on $L^{2}(D)\otimes \C^{N}$ with Dirichlet boundary conditions on $D$. 

\begin{Def}
The Integrated Density of States, or IDS, associated to $H_{A}(\omega)$ is the function from $\R$ to $\R_{+}$, $E\mapsto N(E)$ where $N(E)$ for $E\in \R$ is defined as the following thermodynamical limit : 
\begin{equation}\label{IDSdef}
N(E)=\lim_{L\to +\infty} \frac{1}{|D|} \# \{ \lambda \leq E |\ \lambda \in \sigma (H_{A}^{(D)}(\omega)) \}
\end{equation}
\end{Def}
where $|D|$ is the volume of $D$.

Here we have a double problem of existence in the expression (\ref{IDSdef}). First we have to prove that the cardinal $\# \{ \lambda \leq E |\ \lambda \in \sigma (H_{A}^{(D)}(\omega)) \}$ is finite for each fixed $E$ and then we have to show the existence of the limit. The answer to each one of these problems relies on the existence of an $L^{2}$-kernel for the one-parameter semigroup $(\ee^{-tH_{A}^{(D)}(\omega)})_{t>0}$.

\subsection{A matrix-valued Feynman-Kac formula}

We will first present a matrix-valued Feynman-Kac formula for the one-parameter semigroup $(\ee^{-tH_{A}(\omega)})_{t>0}$ due to Boulton and Restuccia (\cite{boulton}). We will then deduce a Feynman-Kac formula for $(\ee^{-tH_{A}^{(D)}(\omega)})_{t>0}$.
\vskip 2mm

Let $\mathsf{W}=C(\R_{+},\R)$ be the space of continuous functions from $\R_{+}$ to $\R$. For every $t\geq 0$ we consider the coordinate function : 
\begin{align*}
X_{t}\colon\mathsf{W}& \longrightarrow \R \\
\mathsf{w} & \longmapsto X_{t}(\mathsf{w})=\mathsf{w}(t) 
\end{align*}
Let $\mathcal{W}$ be the smallest $\sigma$-algebra on $\mathsf{W}$ for which all the applications $X_{t}$ are measurable. For $s,t\geq 0$ and $x,y\in \R^{d}$ we denote by $W_{s,x,t,y}$ the conditional Wiener measure, defined on $(\mathsf{W},\mathcal{W})$, associated to the Brownian motion starting from $x$ at the time $s$ and arriving on $y$ at the time $t$. We also denote by $\E_{s,x,t,y}$ the expectancy associated to the measure $W_{s,x,t,y}$. For a construction of such conditional Wiener measure and for a construction of the path integral associated to, we refer to \cite{Roe94}, chapter $2$.

We now study the one-parameter semigroup $(\ee^{-tH_{A}(\omega)})_{t>0}$. We fix $t>0$ and $\omega \in \Omega$. By the Lie-Trotter formula we have : 
\begin{equation}\label{lietrotter}
\forall f \in L^{2}(\R^{d})\otimes \C^{N},\ \ee^{-tH_{A}(\omega)}f=\lim_{n\to +\infty} \left( \ee^{-(-\Delta_{d}\otimes I_{N})\frac{t}{n}}\ee^{-V_{\omega}\frac{t}{n}}\right)^{n}f
\end{equation}

For a fixed $n\in \N$, we can use corollary $3.1.2$, p$47$ in \cite{GJ87} to get that the operator : 
$$\left( \ee^{-(-\Delta_{d}\otimes I_{N})\frac{t}{n}}\ee^{-V_{\omega}\frac{t}{n}} \right)^{n}$$
has an integral kernel given by the following path integral : 
\begin{equation}\label{fkkernel}
\int \prod_{j=1}^{n} \ee^{-(\frac{jt}{n}).V_{\omega}(\mathsf{w}(\frac{jt}{n}))}~ \mathrm{d} W_{0,x,t,y}(\mathsf{w})
\end{equation}
But when $n$ tends to infinity we find, by definition of the time-ordered exponential (see \cite{DF79}) : 
\begin{equation}\label{dyson}
\lim_{n \to +\infty} \prod_{j=1}^{n} \ee^{-(\frac{jt}{n}).V_{\omega}(\mathsf{w}(\frac{jt}{n}))} = \exp_{\mathrm{ord}} \left( -\int_{0}^{t} V_{\omega}(\mathsf{w}(s))~ \mathrm{d}s  \right)
\end{equation}
Then by Lebesgue's dominated convergence theorem, we have that : 
\begin{equation}
\forall f \in L^{2}(\R^{d})\otimes \C^{N},\ \forall x \in \R^{d},\ \ee^{-tH_{A}(\omega)}f(x)=\int_{\R^{d}} K_{t}(x,y)f(y)\dd x
\end{equation}
where : 
\begin{equation}\label{noyaueH}
\forall x,y \in \R^{d},\ \forall t>0,\ K_{t}(x,y)=\int \exp_{\mathrm{ord}} \left( -\int_{0}^{t} V_{\omega}(\mathsf{w}(s))~ \mathrm{d}s \right) ~ \mathrm{d} W_{0,x,t,y}(\mathsf{w})
\end{equation}
So we have just proven that $\ee^{-tH_{A}(\omega)}$ has an integral kernel, $K_{t}(x,y)$. Let us see how to deduce from this integral kernel, the existence of an integral kernel for $\ee^{-tH_{A}^{(D)}(\omega)}$. We denote by $T_{D}(\mathsf{w})$ the time of the first exit from $D$ of the path $\mathsf{w} \in \mathsf{W}$ : 
\begin{equation}
T_{D}(\mathsf{w})=\inf \{ t>0,\ X_{t}(\mathsf{w})\notin D \}
\end{equation}
Then the fact that we used Dirichlet boundary conditions to define $H_{A}^{(D)}(\omega)$ allows us to use results on killed Brownian motions (see \cite{Kni81}) which leads to the following formula : 
$$\forall t>0,\ \forall f \in L^{2}(\R^{d})\otimes \C^{N},\ \forall x \in \R^{d},\ \ee^{-tH_{A}^{(D)}(\omega)}f(x)=\qquad \qquad \qquad \qquad \qquad $$
{\footnotesize \begin{equation}\label{killbrownian}
\frac{1}{\sqrt{2\pi t}} \int_{\R^{d}} \int \chi_{\{ t<T(D)(\mathsf{w}) \}} (\mathsf{w}) \exp_{\mathrm{ord}} \left( -\int_{0}^{t} V_{\omega}(X_{s}(\mathsf{w}))~ \mathrm{d}s  \right) ~ \mathrm{d} W_{0,x,t,y}(\mathsf{w})~ \ee^{-\frac{|x-y|^{2}}{2t}}f(y)\dd y 
\end{equation}}
So we have the following proposition : 
\begin{prop}
For every $t>0$, $\ee^{-tH_{A}^{(D)}(\omega)}$ has an integral kernel given by the formula : 
$$\forall x,y \in \R^{d},\ \forall t>0,\ K_{t}^{(D)}(x,y)=\qquad \qquad \qquad \qquad \qquad \qquad \qquad \qquad \qquad $$
{\small \begin{equation}\label{noyauintK}
\frac{1}{\sqrt{2\pi t}}\left( \int \chi_{\{ t<T(D)(\mathsf{w}) \}} (\mathsf{w}) \exp_{\mathrm{ord}} \left( -\int_{0}^{t} V_{\omega}(X_{s}(\mathsf{w}))~ \mathrm{d}s  \right) ~ \mathrm{d} W_{0,x,t,y}(\mathsf{w})~ \ee^{-\frac{|x-y|^{2}}{2t}} \right)
\end{equation} }
And $K_{t}^{(D)}$ is in $L^{2}(D^{2})\otimes \mathcal{M}_{\mathrm{N}}(\C)$ for every $t>0$.
\end{prop}

\begin{proof}
The first assertion and the formula (\ref{noyauintK}) come from (\ref{killbrownian}). Then $D$ is a compact domain in $\R^{d}$ and for a fixed $t>0$, $(x,y)\mapsto K_{t}^{(D)}(x,y)$ is continuous. As in (\ref{noyauintK}), $t$ is bounded by $T_{D}(\mathsf{w})$, we have that $K_{t}^{(D)}$ is in $L^{2}(D^{2})\otimes \mathcal{M}_{\mathrm{N}}(\C)$ as it is a bounded continuous function on $D^{2}$.
\end{proof}

This proposition will be the main ingredient to prove the existence of the IDS associated to $H_{A}(\omega)$.

\subsection{Existence of the IDS}

From Proposition \ref{noyauintK}, we deduce that for every $t>0$, the operator $\ee^{-tH_{A}^{(D)}(\omega)}$ is Hilbert-Schmidt on $L^{2}(D)\otimes \C^{N}$. Thus, its spectrum is of the form :
$$\{ \ee^{-t\lambda_{j}^{(D)}(\omega)},\ j\geq 0 \}$$
where $(\lambda_{j}^{(D)}(\omega))_{j\geq 0}$ is an increasing sequence of real numbers, bounded from below and tending to $+\infty$. This sequence is the spectrum of $H_{A}^{(D)}(\omega)$. In particular, for a fixed $E\in \R$ : 
$$\# \{ \lambda \leq E\ |\ \lambda \in \sigma (H_{A}^{(D)}(\omega))\} = \# \{ \lambda_{j}^{(D)}(\omega) \leq E \} < +\infty$$
This answers the first part of the problem of existence of $N(E)$. It remains to prove that the sequence $\frac{1}{|D|} \# \{ \lambda_{j}^{(D)}(\omega) \leq E  \}$ converges to a real number independent of $\omega$: $N(E)$. To that end, we introduce the counting measure of the eigenvalues of $H_{A}^{(D)}(\omega)$ : 
\begin{equation}\label{countmeasure}
\mathfrak{n}_{D,\omega}=\frac{1}{|D|} \sum_{j \geq 0} \delta_{\lambda_{j}^{(D)}(\omega)}
\end{equation}
where $\delta_{\lambda_{j}^{(D)}(\omega)}$ is the Dirac measure at $\lambda_{j}^{(D)}(\omega)$. Then we have : 

\begin{prop}\label{laplaceIDS}
The sequence of measures $(\mathfrak{n}_{D,\omega})_{L\geq 1}$ converges vaguely to a measure $\mathfrak{n}$ independent of $\omega$ as $L$ tends to $+\infty$ for $\mathsf{P}$-almost every $\omega$ in $\Omega$. Moreover, the Laplace transform of this measure $\mathfrak{n}$ is given by: $\forall t>0$,  
\begin{equation}\label{laplacen}
L(\mathfrak{n})(t)=\frac{1}{\sqrt{2\pi t}} \int \int_{\Omega} \mathrm{Tr}_{\C^{N}} \exp_{\mathrm{ord}} \left( -\int_{0}^{t} V_{\omega}(X_{s}(\mathsf{w}))~ \mathrm{d}s \right)~ \mathrm{d}\omega ~\mathrm{d}W_{0,0,t,0}(\mathsf{w})
\end{equation}
\end{prop}

\begin{cor}
For every $E\in \R$, the limit :
$$N(E)=\lim_{L\to +\infty} \frac{1}{|D|} \# \{ \lambda \leq E |\ \lambda \in \sigma (H_{A}^{(D)}(\omega)) \}$$
exists and is $\mathsf{P}$-almost surely independent of $\omega$. The function $E\mapsto N(E)$ is the repartition function of $\mathfrak{n}$ : 
$$\forall E\in \R,\ N(E)=\mathfrak{n}([E,+\infty))$$
\end{cor}

Before proving this proposition, we need to prove a lemma which gives the expression of the trace of an operator with matrix-valued integral kernel. We adapt here a result of Simon proven in \cite{sim05}, thm $3.9$, p.$35$.

\begin{lem}\label{formuletrace}
Let $H$ be a self-adjoint operator acting on $L^{2}(D)\otimes \C^{N}$ where $D\subset \R^{d}$ is a compact set. We assume that for all $t>0$ the operator $\ee^{-tH}$ is class-trace and has a matrix-valued integral kernel $K_{t}$. Then:
$$\mathrm{Tr}(\ee^{-tH}) = \int_{D} \mathrm{Tr}_{\C^{N}} K_{t}(x,x)\dd x $$
where $\mathrm{Tr}_{\C^{N}}$ denotes the usual trace on $N\times N$ matrices.
\end{lem}

\begin{proof}
Let $n\in \N$, $m\in \{0,\ldots,2^{n} \}$ and $k\in \{ 1,\ldots,N\}$. We set:
{\small $$\phi_{n,m,k}(x)=\left\lbrace \begin{array}{ccl}
^{t}(0,\ldots,0,2^{\frac{n}{2}},0,\ldots,0) & \mathrm{if} & \forall i\in \{1,\ldots N\},\  -L.\frac{m-1}{2^{n}}\leq x_{i} < L.\frac{m}{2^{n}} \\
^{t}(0,\ldots,0) & \mathrm{otherwise} & 
\end{array}\right.$$}
where $2^{\frac{n}{2}}$ is at the $k$-th position. Then the family $\{ \phi_{n,m,k} \}_{n\in \N, 0\leq m\leq 2^{n}, 1\leq k \leq N}$ is a Hilbert basis of the Hilbert space $L^{2}(D)\otimes \C^{N}$.

Let $P_{n}$ be the projection on the subspace spanned by the $2^{n}N$ functions $\phi_{n,m,k}$ for $n$ fixed and $m\in \{0,\ldots,2^{n} \}$, $k\in \{ 1,\ldots,N\}$. Then one can construct an Hilbert basis $(\psi_{1},\psi_{2},\ldots)$ of $L^{2}(D)\otimes \C^{N}$ such that : 
$$\forall n\in \N,\ \psi_{1},\ldots,\psi_{2^{n}N} \in \mathrm{Im}~P_{n}$$
Then we have : 
$$\mathrm{Tr}(\ee^{-tH})=\lim_{n\to +\infty}\mathrm{Tr}(P_{n}\ee^{-tH}P_{n}) $$
by \emph{Th} $3.1$, p$31$ in \cite{sim05}. But : 
$$\forall n\in \N,\ \mathrm{Tr}(P_{n}\ee^{-tH}P_{n}) =  \sum_{k=1}^{N} \sum_{m=1}^{2^{n}} (\phi_{n,m,k},\ee^{-tH}\phi_{n,m,k})\qquad \qquad \qquad \qquad$$
\begin{eqnarray}
& = & \sum_{k=1}^{N} \sum_{m=1}^{2^{n}} \int_{D} \int_{D} {^{t}\overline{\phi_{n,m,k}(x)}} K_{t}(x,y)\phi_{n,m,k}(y)\dd x \mathrm{d}y \nonumber \\ 
& = & \sum_{m=1}^{2^{n}} \int \int_{-L.\frac{m-1}{2^{n}}\leq x_{i},y_{i} < L.\frac{m}{2^{n}}} 2^{\frac{n}{2}}.2^{\frac{n}{2}} \underbrace{\left( \sum_{k=1}^{N} (0,\ldots,1,\ldots,0)K_{t}(x,y) ^{t}(0,\ldots,1,\ldots,0) \right)}_{\mathrm{Tr}_{\C^{N}}(K_{t}(x,y))}\dd x \mathrm{d}y \nonumber \\ 
& = & 2^{n} \sum_{m=1}^{2^{n}} \int \int_{-L.\frac{m-1}{2^{n}}\leq x_{i},y_{i} < L.\frac{m}{2^{n}}} \mathrm{Tr}_{\C^{N}}(K_{t}(x,y)) \dd x \mathrm{d}y  \nonumber
\end{eqnarray} 
Then by uniform continuity of $K_{t}$ on the compact set $D^{2}$ : 
{\small $$
\lim_{n\to +\infty} 2^{n} \sum_{m=1}^{2^{n}} \int \int_{-L.\frac{m-1}{2^{n}}\leq x_{i},y_{i} < L.\frac{m}{2^{n}}} \mathrm{Tr}_{\C^{N}}(K_{t}(x,y)) \dd x \mathrm{d}y = \int_{D}  \mathrm{Tr}_{\C^{N}}(K_{t}(x,x)) \dd x $$}
\end{proof}

\begin{proof}[Proposition \ref{laplaceIDS}]
We fix $t>0$. We have : 
{\small 
\begin{eqnarray}
L(\mathfrak{n}_{D,\omega})(t) & = & \int_{\R} \ee^{-Et} \mathfrak{n}_{D,\omega}(E)  \nonumber \\
 & = & \frac{1}{|D|} \sum_{j\geq 0} \ee^{-\lambda_{j}^{(D)}(\omega)t} \nonumber \\
 & = & \frac{1}{|D|} \mathrm{Tr}(\ee^{-tH_{A}^{(D)}(\omega)}) \nonumber \\
 & = & \frac{1}{|D|} \int_{D}  \mathrm{Tr}_{\C^{N}}(K_{t}(x,x)) \dd x \nonumber \\
 & = & \frac{1}{|D|} \frac{1}{\sqrt{2\pi t}} \int_{D} \int \chi_{\{ t<T_{D}(\mathsf{w})\} }(\mathsf{w}) \mathrm{Tr}_{\C^{N}} \exp_{\mathrm{ord}} \left( -\int_{0}^{t} V_{\omega}(X_{s}(\mathsf{w}))~ \mathrm{d}s \right)~ \mathrm{d}W_{0,x,t,x}(\mathsf{w})\dd x \nonumber
\end{eqnarray}}
by (\ref{noyauintK}). We set: 
{\small \begin{equation}\label{eqnAlaplace}
A_{D}=\frac{1}{|D|} \frac{1}{\sqrt{2\pi t}} \int_{D} \int \mathrm{Tr}_{\C^{N}} \exp_{\mathrm{ord}} \left( -\int_{0}^{t} V_{\omega}(X_{s}(\mathsf{w}))~ \mathrm{d}s \right)~ \mathrm{d}W_{0,x,t,x}(\mathsf{w})\dd x
\end{equation}}
and: 
{\small \begin{equation}\label{eqnBlaplace}
B_{D}=\frac{1}{|D|} \frac{1}{\sqrt{2\pi t}} \int_{D} \int \chi_{\{ t\geq T_{D}(\mathsf{w})\} }(\mathsf{w}) \mathrm{Tr}_{\C^{N}} \exp_{\mathrm{ord}} \left( -\int_{0}^{t} V_{\omega}(X_{s}(\mathsf{w}))~ \mathrm{d}s \right)~ \mathrm{d}W_{0,x,t,x}(\mathsf{w})\dd x
\end{equation}}
\vskip 2mm

Using Birkhoff's theorem when $L\to +\infty$ in $A_{D}$, we get : 
{\small \begin{equation}\label{cveqnAlaplace}
\lim_{L\to +\infty} A_{D} = \frac{1}{\sqrt{2\pi t}} \int \int_{\Omega} \mathrm{Tr}_{\C^{N}} \exp_{\mathrm{ord}} \left( -\int_{0}^{t} V_{\omega}(X_{s}(\mathsf{w})) ~ \mathrm{d}s \right)~\mathrm{d}\omega ~ \mathrm{d}W_{0,0,t,0}(\mathsf{w})
\end{equation}}
Let $\mathfrak{n}$ be the measure on $\R$ (with the Borel $\sigma$-algebra) such that:
\begin{equation}\label{defn}
L(\mathfrak{n})(t)=\frac{1}{\sqrt{2\pi t}} \int \int_{\Omega} \mathrm{Tr}_{\C^{N}} \exp_{\mathrm{ord}} \left( -\int_{0}^{t} V_{\omega}(X_{s}(\mathsf{w}))  ~ \mathrm{d}s \right) ~\mathrm{d}\omega ~\mathrm{d}W_{0,0,t,0}(\mathsf{w})
\end{equation}  
To prove that $\mathfrak{n}_{D,\omega}$ converges vaguely to $\mathfrak{n}$ as $L$ tends to infinity, it remains to prove that $B_{D}\to 0$ and that the convergence of $A_{D}$ and $B_{D}$ happens on a set $\Omega_{1}$ independent of $t$ and of measure $1$. Actually, for the rest of the proof, we can refer to the proof of Carmona in \cite{carmonastflour}, \emph{Th.}$V1$, p.$66-67$. Indeed, as $V_{\omega}$ is uniformly bounded on $\R$ in $x$ and in  $\omega$, the function: 
\begin{equation}\label{ramenecarmona}
\begin{array}{ccl}
\Omega \times \mathsf{W} & \to & \C \\
(\omega, \mathsf{w}) & \to & \mathrm{Tr}_{\C^{N}} \exp_{\mathrm{ord}} \left( -\int_{0}^{t} V_{\omega}(X_{s}(\mathsf{w})) \right)
\end{array}
\end{equation}
for $t>0$ fixed is in every $L^{r}(\Omega \times \mathsf{W},\mathsf{P}\otimes W_{0,0})$ for all $r>1$. Here $W_{0,0}$ is the Wiener measure defined on $(\mathsf{W},\mathcal{W})$ associated to the Brownian motion starting from $0$ at time $0$. Thus, function (\ref{ramenecarmona}) has the same properties as the function: 
\begin{equation}\label{ramenecarmona2}
\begin{array}{ccl}
\Omega \times \mathsf{W} & \to & \C \\
(\omega, \mathsf{w}) & \to & \exp \left( -\int_{0}^{t} q^{-}(X_{s}(\mathsf{w}),\omega) \right)
\end{array}
\end{equation}
in \cite{carmonastflour}, \emph{Th.}$V1$. Then one can rewrite the end of the proof of \cite{carmonastflour} by changing (\ref{ramenecarmona2}) by (\ref{ramenecarmona}).
\end{proof}
\vskip 2mm

\noindent \emph{Remark: } In the proof of Proposition \ref{laplaceIDS}, we did not verify that the limit measure $\mathfrak{n}$ does not depend on the choice of boundary conditions for $H_{A}^{(D)}(\omega)$. This choice appears in formula (\ref{killbrownian}) by introducing the characteristic function $\chi_{\{ t<T_{D}(\omega)\} }$ corresponding to a killed Brownian motion. If by example we had chosen Neumann boundary conditions instead of Dirichlet boundary conditions we should had to change this characteristic function to make it correspond to a reflected Brownian motion (see \cite{Kni81}, chapter $4$). The rest of the proofs is unchanged and the expression of $\mathfrak{n}$ does not depend on $\chi_{\{ t<T_{D}(\omega)\} }$.
\vskip 2mm

We finish this section by proving a formula which relates the measure $\mathfrak{n}$ to the spectral measure associated to the self-adjoint operator $H_{A}(\omega)$. This spectral measure will be denoted by: $E_{H_{A}(\omega)}$.

\begin{prop}\label{idsEtrace}
Let $f$ be a continuous, positive, compactly supported function on $\R^{d}$, such that $||f||_{L^{2}(\R^{d})}=1$. We denote by $M_{f}$ the maximal multiplication operator by $f$. Then for every bounded Borel set $B$ of $\R$, the operator $M_{f}E_{H_{A}(\omega)}(B)M_{f}$ is trace-class $\mathsf{P}$-almost surely in $\omega$ and: 
\begin{equation}\label{idsEtraceeq}
\mathfrak{n}(B)=\E(\mathrm{Tr}(M_{f}E_{H_{A}(\omega)}(B)M_{f}))
\end{equation}
where $\E$ is the expectancy associated to the probability measure $\mathsf{P}$.
\end{prop} 

\begin{proof}
If $B\subset \R$ is a bounded Borel set of $\R$, then there exist strictly positives constants $C$ and $t$ such that : 
\begin{equation}\label{ineqborel}
\forall x\in \R,\ \chi_{B}(x)\leq C\ee^{-tx}
\end{equation}
Let $\{ f_{k} \}_{k\geq 1}$ be a Hilbert basis of $L^{2}(\R^{d})\otimes \C^{N}$. Let $f$ be a positive, continuous, compactly supported function on $\R^{d}$, such that $||f||_{L^{2}(\R^{d})}=1$. Then:
{\small $$\E\left( \sum_{k\geq 1} <(M_{f}E_{H_{A}(\omega)}(B)M_{f})f_{k},f_{k}>\right)  \leq  C\E\left( \sum_{k\geq 1} <\ee^{-tH_{A}(\omega)}(ff_{k}),(ff_{k})>\right) 
$$}
by the spectral theorem applicated to $\chi_{B}$, the inequality (\ref{ineqborel}) and the fact that $M_{f}$ is self-adjoint as $f$ is real-valued. But: 
$$\E\left( \sum_{k\geq 1} <\ee^{-tH_{A}(\omega)}(ff_{k}),(ff_{k})>\right)=\E\left( \mathrm{Tr}(M_{f}\ee^{-tH_{A}(\omega)}M_{f}) \right)$$
Let $L$ be large enough for $D=[-L,L]^{d}$ to contain the support of $f$. Then using lemma \ref{formuletrace}: 
\begin{eqnarray}
\E\left( \mathrm{Tr}(M_{f}\ee^{-tH_{A}(\omega)}M_{f}) \right) &=& \E \left( \int_{\supp f} f(x)^{2} \mathrm{Tr}_{\C^{N}} K_{t}(x,x) ~\mathrm{d}x\right) \nonumber \\
&=& \E \left( \int_{\R} f(x)^{2} \mathrm{Tr}_{\C^{N}} K_{t}(x,x) ~\mathrm{d}x\right) \label{215}
\end{eqnarray}
with $K_{t}$ given by (\ref{noyaueH}). Then, using the $\R^{d}$-ergodicity of $H_{A}(\omega)$ at the second equality:
$$\E \left( \int_{\R^{d}} f(x)^{2} \mathrm{Tr}_{\C^{N}} K_{t}(x,x)~\mathrm{d}x\right) \qquad \qquad \qquad \qquad \qquad \qquad \qquad \qquad \qquad $$
\begin{eqnarray}
 & = &  \frac{1}{\sqrt{2\pi t}}\E \left( \int_{\R^{d}} f(x)^2 \int \mathrm{Tr}_{\C^{N}} \exp_{\mathrm{ord}} \left( -\int_{0}^{t} V_{\omega}(\mathsf{w}(s))~ \mathrm{d}s \right) ~ \mathrm{d} W_{0,x,t,x}(\mathsf{w})~\mathrm{d}x\right) \nonumber \\
 & = &  \frac{1}{\sqrt{2\pi t}}\E \left( \int_{\R^{d}} f(x)^2  \int \mathrm{Tr}_{\C^{N}} \exp_{\mathrm{ord}} \left( -\int_{0}^{t} V_{\omega}(x+\mathsf{w}(s))~ \mathrm{d}s \right) ~ \mathrm{d} W_{0,0,t,0}(\mathsf{w})~\mathrm{d}x\right) \nonumber \\
 & = &  \frac{1}{\sqrt{2\pi t}}\E \left( \int_{\R^{d}} f(x)^2  \int \mathrm{Tr}_{\C^{N}} \exp_{\mathrm{ord}} \left( -\int_{0}^{t} V_{\omega}(\mathsf{w}(s))~ \mathrm{d}s \right) ~ \mathrm{d} W_{0,0,t,0}(\mathsf{w})~\mathrm{d}x\right) \nonumber \\
 & = &  \frac{1}{\sqrt{2\pi t}}\E \left(  \int \mathrm{Tr}_{\C^{N}} \exp_{\mathrm{ord}} \left( -\int_{0}^{t} V_{\omega}(\mathsf{w}(s))~ \mathrm{d}s \right) ~ \mathrm{d} W_{0,0,t,0}(\mathsf{w})\right) \label{216} 
\end{eqnarray} 
And this last expectancy is finite by Proposition \ref{laplaceIDS}. So we have proved that:
$$\E\left( \sum_{k\geq 1} <(M_{f}E_{H_{A}(\omega)}(B)M_{f})f_{k},f_{k}>\right) \qquad \qquad \qquad \qquad \qquad \qquad \qquad \qquad $$
\begin{equation}
\leq C \frac{1}{\sqrt{2\pi t}}\E \left(  \int \mathrm{Tr}_{\C^{N}} \exp_{\mathrm{ord}} \left( -\int_{0}^{t} V_{\omega}(\mathsf{w}(s))~ \mathrm{d}s \right) ~ \mathrm{d} W_{0,0,t,0}(\mathsf{w})\right) <+\infty
\end{equation} 
which means that the operator $M_{f}E_{H_{A}(\omega)}(B)M_{f}$ is trace class $\mathsf{P}$-almost surely on $\omega \in \Omega$. It also proves that $B\mapsto \E( \mathrm{Tr}(M_{f}E_{H_{A}(\omega)}(B)M_{f}))$ defines a Radon measure on $\R$ whose Laplace transform is:
\begin{equation}
L(\E(\mathrm{Tr}(M_{f}E_{H_{A}(\omega)}(.)M_{f})))(t)=\E(\mathrm{Tr}(M_{f}\ee^{-tH_{A}(\omega)}M_{f}))=L(\mathfrak{n})(t)
\end{equation}
by (\ref{216}), (\ref{215}) and (\ref{laplacen}). By injectivity of the Laplace transform, we have that for every bounded Borel set $B\subset \R$:
$$\mathfrak{n}(B)=\E(\mathrm{Tr}(M_{f}E_{H_{A}(\omega)}(B)M_{f}))$$ 
\end{proof}
\vskip 2mm

All the results of this section were valid for $H_{A}(\omega)$ acting on $L^{2}(\R^{d})\otimes \C^{N}$ for every $d$ and every $N$. In the next sections, we will restrict our presentation to the case of $d=1$ and $N$ arbitrary, $N\geq 1$. It will allow us to introduce the Lyapounov exponents associated to $H_{A}(\omega)$.

We want to study the regularity of the function $E\mapsto N(E)$. As an increasing function we already know that it has left and right limits at each point of the real line. We will actually prove that the IDS is locally H\"older continuous. To prove this, we will prove the same regularity property for the Lyapounov exponents associated to $H_{A}(\omega)$ and show that the IDS and the Lyapounov exponents are related to each other trough an harmonic analysis formula, a Thouless formula.

\section{Lyapounov exponents}\label{section3}

\subsection{Definition and integral representation}

We start with a review of some results about Lyapounov exponents. These results holds for general sequences of independent and identically distributed (\emph{i.i.d.}) random symplectic matrices. Let $N$ be a positive integer. Let $\mathrm{Sp}_{N}(\R)$ denote the group of $2N\times 2N$ real symplectic matrices. It is the subgroup of $\mathrm{GL}_{2N}(\R)$ of matrices $M$ satisfying $$^tMJM=J,$$ 
where $J$ is the matrix of order $2N$ defined by $J=\bigl(\begin{smallmatrix}
0 & -I_{N} \\
I_{N} & 0
\end{smallmatrix}\bigr)$. 

\begin{Def}
Let $(A_{n}^{\omega})_{n\in \N}$ be a sequence of i.i.d.\ random
matrices in $\mathrm{Sp}_{N}(\R)$ with $$\mathbb{E}(\log^{+}
||A_{0}^{\omega}||) <\infty.$$ The Lyapunov exponents
$\gamma_{1},\ldots,\gamma_{2N}$ associated with
$(A_{n}^{\omega})_{n\in \N}$ are defined inductively by
\begin{equation}\label{lyapdef}
\sum_{i=1}^{p} \gamma_{i} = \lim_{n \to \infty} \frac{1}{n}
\mathbb{E}(\log ||\wedge^{p} (A_{n-1}^{\omega}\ldots A_{0}^{\omega})
||)
\end{equation}
for $p\in \{1,\ldots,N\}$.
\end{Def}

Here, $\wedge^{p} M$ denotes the $p$-th exterior power of the matrix $M$, acting on the $p$-th exterior power of $\R^{2N}$. One has $\gamma_{1}\geq \ldots \geq \gamma_{2N}$. Moreover, the random matrices $(A_{n})_{n\in \N}$ being symplectic, we have the symmetry property $\gamma_{2N-i+1}= -\gamma_{i},\ \forall i \in \{1,\ldots,N\}$ (see \cite{bougerol} p.$89$, Prop $3.2$).

Let $\mu$ be a probability measure on $\mathrm{Sp}_{N}(\R)$. We denote by $G_{\mu}$ the smallest closed subgroup of $\mathrm{Sp}_{N}(\R)$ which contains the topological support of
$\mu$, $\mathrm{supp}\ \mu$. We also define for every $p\in \{ 1,\ldots ,N\}$, the $p$-Lagrangian submanifold $L_{p}$ of $\R^{2N}$, as the subspace of $\wedge^{p}\R^{2N}$ spanned by $\{Me_1 \wedge \ldots \wedge Me_p \ |
\ M\in \mathrm{Sp}_{N}(\R) \}$, where $(e_{1},\ldots,e_{2N})$ is the canonical basis of $\R^{2N}$. 

We can now give a generalization of F\"urstenberg's theorem for $N>1$. For the definitions of $L_p$-strong irreducibility and $p$-contractivity we refer to \cite{bougerol}, definitions
$A.IV.3.3$ and $A.IV.1.1$, respectively.

\begin{prop}\label{lyapsepmain}
Let $(A_{n}^{\omega})_{n\in \N}$ be a sequence of i.i.d.\ random
symplectic matrices of order $2N$ and $p$ be an integer, $1\leq p\leq N$. Let $\mu$ be 
the common distribution of the $A_{n}^{\omega}$. If 
\begin{enumerate}[\rm(a)]
\item
$G_{\mu}$ is $p$-contracting and $L_p$-strongly irreducible,
\item
$\mathbb{E}(\log\Vert A_{0}^{\omega}\Vert)<\infty$, 
\end{enumerate}
then the following holds:
\begin{enumerate}[\rm(i)]
\item
$\gamma_{p} > \gamma_{p+1}$
\item
For any non zero $x$ in $L_p$:
\[
\lim_{n\to \infty} \frac{1}{n}\mathbb{E}\bigl(\log\Vert(\wedge^{p} A_{n-1}^{\omega}\ldots A_{0}^{\omega})x\Vert\bigr)=\sum_{i=1}^{p} \gamma_i\,.
\]
\item
There exists a unique $\mu$-invariant probability measure $\nu_{p}$ on $\mathbb{P}(L_{p})= \{ \bar{x} \in \mathbb{P}(\wedge^{p} \R^{2N})\ |\ x\in \ L_{p} \}$ such that: 
\[
\sum_{i=1}^{p} \gamma_i = \int_{\Sp_{N}(\R)\times \mathbb{P}(L_{p})} \log \frac{||\wedge^{p} M x||}{||x||}\dd\mu(M)\dd\nu_{p} (\bar{x})
\] 
\end{enumerate}
\end{prop}

\begin{proof}
This is Proposition~3.4 of \cite{bougerol}.
\end{proof}

It remains to define the Lyapounov exponents associated to the operator $H_{A}(\omega)$ for $d=1$ and $N\geq 1$. For $E\in \R$ we can consider the second order differential system : \begin{equation}\label{system2nd}
H_{A}(\omega) u = Eu \Leftrightarrow -u''+ V_{\omega} u = Eu
\end{equation}
with $u=(u_1 , \ldots , u_N) \in \C^{N}$. We introduce the transfer matrix $A_{n}^{\omega}(E)$ from $n$ to $n+1$, defined by the relation:
\begin{equation}\label{deftransfer}
\left( \begin{array}{c}
u(n+1,E) \\
u'(n+1,E) 
\end{array} \right) = A_{n}^{\omega}(E) \left( \begin{array}{c}
u(n,E) \\
u'(n,E) 
\end{array} \right)
\end{equation}
Then one can verify that $(A_{n}^{\omega}(E))_{n\in \N}$ is a sequence of \emph{i.i.d.} random symplectic matrices because the system (\ref{system2nd}) is Hamiltonian. So we can define the Lyapounov exponents associated to the operator $H_{A}(\omega)$ as the Lyapounov exponents of the sequence of transfer matrices $(A_{n}^{\omega}(E))_{n\in \N}$. Since the transfer matrices depend on a real parameter $E$, so will the Lyapounov exponents of $H_{A}(\omega)$ and so do the measure $\mu_{E}$ (the common law of the $A_{n}^{\omega}(E)$), the group $G_{\mu_{E}}$ and the $\mu_{E}$-invariant probability measure $\nu_{p,E}$ of proposition \ref{lyapsepmain}.

\subsection{Regularity of the Lyapounov exponents}

We want to study the regularity of the function $E\mapsto \gamma_{p}(E)$ for $p\in \{1,\ldots,N\}$. According to the integral representation obtained at Proposition \ref{lyapsepmain}, we have to understand the regularity of $E\mapsto \nu_{p,E}$ for any $p$ in $\{1,\ldots,N\}$ and to control the term $||\wedge^{p} M||$ in the integral, which depend on $E$ as $\mu_{E}$ depends on $E$. We will now give a general theorem for the regularity of the Lyapounov exponents of sequences of \emph{i.i.d.} random symplectic matrices depending on a real parameter.

\begin{thm}\label{holderlyapgen}
Let $(A_{n}^{\omega}(E))_{n\in \N}$ be a sequence of \emph{i.i.d.} random symplectic matrices depending on a real parameter E. Let $\mu_{E}$ be the common distribution of the $A_{n}^{\omega}(E)$. We fix a compact interval $I$ in $\R$ and we assume that for $E\in I$ we have:
\begin{enumerate}[\rm(i)]
\item $G_{\mu_{E}}$ is $p$-contracting and $L_p$-strongly irreducible for every $p\in \{1,\ldots ,N\}$.
\item
There exist $C_{1}>0$, $C_{2}>0$ independent of $n,\omega,E$ such that for every $p\in \{1,\ldots ,N\}$:
\begin{equation}\label{estim1}
||\wedge^{p} A_{n}^{\omega}(E)||^{2} \leq \exp( pC_{1} + p|E| +p) \leq C_{2}
\end{equation}
\item
There exists $C_{3}>0$ independent of $n,\omega,E$ such that for every $E,E'\in I$ and every $p\in \{1,\ldots ,N\}$: 
\begin{equation}\label{estim2}
||\wedge^{p} A_{n}^{\omega}(E)-\wedge^{p} A_{n}^{\omega}(E')||\leq C_{3} |E-E'|
\end{equation}
\end{enumerate}
\vskip 2mm

\noindent Then there exist two real numbers $\alpha >0$ and $0<C<+\infty$ such that:
$$\forall p\in \{1,\ldots N\},\ \forall E,E' \in I,\ |\gamma_{p}(E)-\gamma_{p}(E')| \leq C |E-E'|^{\alpha}$$
\end{thm}

\begin{proof}
The methods to prove this theorem can be found in \cite{carmona}, chapter V. In this reference this regularity result is written for transfer matrices associated to matrix-valued discrete Schr\"odinger operators. But this restriction to discrete operators only concerns the estimates (\ref{estim1}) and (\ref{estim2}). They are obviously verified in the case of transfer matrices of discrete Schr\"odinger operators as it is explained in \cite{carmona}, p.279. For a presentation using estimates (\ref{estim1}) and (\ref{estim2}), one can read \cite{DSS02} where it is done in the case of transfer matrices associated to scalar-valued continuous Schr\"odinger operators.

The main steps of the proof are the following. First we prove continuity of the Lyapounov exponents on $I$ by proving continuity of the function:
\begin{align*}
\Phi_{p,E}\colon
I\times \P(L_{p}) & \longrightarrow  \R \\
(E, \bar{x}) & \longmapsto \Phi_{p,E}(\bar{x})=\E \left( \log \frac{||(\wedge^{p} A_{n}^{\omega}(E))x||}{||x||} \right)
\end{align*} 
for every $p\in \{1,\ldots N\}$. We only use estimates (\ref{estim1}) and (\ref{estim2}) to prove this continuity. Then we prove weak continuity of the function $E\mapsto \nu_{p,E}$ using Banach-Alaoglu theorem and the unicity of the $\mu_{E}$-invariant measure $\nu_{p,E}$ as stated in point $(iii)$ of proposition \ref{lyapsepmain}. Combining these two continuity properties and noting that: 
$$\gamma_{1}(E)+\ldots +\gamma_{p}(E)=\nu_{p,E}(\Phi_{p,E})$$
we get the continuity of the Lyapounov exponents.

To prove the H\"older continuity of the Lyapounov exponents we need a result on negative cocyles as stated in \cite{carmona}, Proposition IV 3.5, p.187. We also need estimates on Laplace operators on H\"older spaces like Proposition V 4.13, p.277 in \cite{carmona} which relies on estimates (\ref{estim1}) and (\ref{estim2}). Finally using the decomposition given in Proposition IV 3.12, p.192 in \cite{carmona} one can prove  the H\"older continuity of $E\mapsto \nu_{p,E}$ on $I$.

For a complete presentation of this proof in the case of transfer matrices for continuous matrix-valued Schr\"odinger operators, with proofs showing the role of the $p$-th exterior powers, we refer to \cite{boumazathese}, chapter 6.
\end{proof}

We will now use this general result to prove the following theorem:

\begin{thm}\label{holderlyap}
Let $I$ be a compact interval in $\R$. We assume that the potential $V_{\omega}$ in $H_{A}(\omega)$ for $d=1$ and $N\geq 1$ is such that the group $G_{\mu_{E}}$ associated to the transfer matrices of $H_{A}(\omega)$ is $p$-contracting and $L_{p}$-strongly irreducible for every $p\in \{1,\ldots ,N\}$ and all $E\in I$. Then the Lyapounov exponents associated to $H_{A}(\omega)$ are H\"older continuous on $I$, \emph{i.e}, there exist two real numbers $\alpha >0$ and $0<C<+\infty$ such that:
$$\forall p\in \{1,\ldots N\},\ \forall E,E' \in I,\ |\gamma_{p}(E)-\gamma_{p}(E')| \leq C |E-E'|^{\alpha}$$
\end{thm}

According to theorem \ref{holderlyapgen} we only have to show that the transfer matrices $A_{n}^{\omega}(E)$ associated to $H_{A}(\omega)$ verify estimates (\ref{estim1}) and (\ref{estim2}). They already verify point $(i)$ of theorem \ref{holderlyapgen} by assumption. Before proving (\ref{estim1}) and (\ref{estim2}) we will give two lemmas which are the analog for matrix-valued operators of lemmas A.1 and A.2 in \cite{DSS02}.

\begin{lem}\label{lemmaapriori1}
Let $V$ be a matrix-valued function in $L^{1}_{\mathrm{loc}}(\R,M_{N}(\R))$ and $u$ a solution of $-u''+Vu=0$. Then for all $x,y\in \R$ : 
$$||u(x)||^{2}+||u'(x)||^{2} \leq (||u(y)||^{2}+||u'(y)||^{2})\exp \left(\int_{\min(x,y)}^{\max(x,y)} ||V(t)+1||dt \right)$$ 
\end{lem}

\begin{proof}
Let $R(t)=||u(t)||^{2}+||u'(t)||^{2}$. We have : 
\begin{eqnarray}
R'(t) & = & <u(t),u'(t)>+<u'(t),u(t)>+<u''(t),u'(t)>+<u'(t),u''(t)> \nonumber \\
 & = & 2\mathrm{Re}(<u(t),u'(t)>)+2\mathrm{Re}(<u'(t),V(t)u(t)>) \nonumber \\
 & = & 2\mathrm{Re}(<u'(t),(V(t)+1)u(t)>) \nonumber \\
 & \leq & 2 \mathrm{Re}(||u'(t)||\ ||V(t)+1||\ ||u(t)||) \nonumber \\
 & \leq & 2||V(t)+1||\left( \frac{||u(t)||^{2}+||u'(t)||^{2}}{2}\right) \nonumber \\
 & = & ||V(t)+1||R(t) \nonumber
\end{eqnarray}

We have used the Cauchy-Schwarz inequality and the arithmetico-geometric inequality. Finally, we have the inequality : 
$$R'(t)\leq ||V(t)+1||R(t)$$
Which by integration gives us the expected inequality.
\end{proof}

\begin{lem}\label{lemmaapriori2}
For $i=1,2$ let $V_{i}\in L^{1}_{\mathrm{loc}}(\R,M_{N}(\R))$ and $u_{i}$ a solution of $-u''+V_{i}u=0$ such that : 
$$\exists y \in \R,\ u_{1}(y)=u_{2}(y)\ \mathrm{and} \ u_{1}'(y)=u_{2}'(y)$$
Then, for every $x\in \R$ :
$$\left( ||u_{1}(x)-u_{2}(x)||^{2}+||u_{1}'(x)-u_{2}'(x)||^{2} \right)^{\frac{1}{2}}\leq \left(  ||u_{1}(y)||^{2}+||u_{1}'(y)||^{2} \right)^{\frac{1}{2}} \times \qquad $$
 $$ \qquad \qquad  \exp \left( \int_{\min(x,y)}^{\max(x,y)} ||V_{1}(t)||+||V_{2}(t)||+2\; dt\right) \times \int_{\min(x,y)}^{\max(x,y)} ||V_{1}(t)-V_{2}(t)||dt$$

\end{lem}

\begin{proof}
Without loss of generality we can assume that $y\leq x$. We have, because of the assumptions made on the solutions $u_{1}$ and $u_{2}$ : 
{\footnotesize $$\left( \begin{array}{c}
u_{1}(x)-u_{2}(x) \\
u_{1}'(x)-u_{2}'(x)
\end{array}\right) = \int_{x}^{y} \left( \begin{array}{c}
0 \\
(V_{1}(t)-V_{2}(t))u_{1}(t)
\end{array}\right) dt + \int_{x}^{y} \left( \begin{array}{cc}
0  & I\\
V_{2}(t) & 0
\end{array}\right)\left( \begin{array}{c}
u_{1}(t)-u_{2}(t) \\
u_{1}'(t)-u_{2}'(t)
\end{array}\right) dt$$}
We take the norm of the two sides of the equality : 
{\footnotesize $$\left|\left|\left( \begin{array}{c}
u_{1}(x)-u_{2}(x) \\
u_{1}'(x)-u_{2}'(x)
\end{array}\right)\right|\right| \leq \int_{x}^{y} ||V_{1}(t)-V_{2}(t)||\ ||u_{1}(t)||dt + \int_{x}^{y} (||V_{2}(t)||+1)\left|\left|\left( \begin{array}{c}
u_{1}(t)-u_{2}(t) \\
u_{1}'(t)-u_{2}'(t)
\end{array}\right)\right|\right|dt$$}
Then by Gronwall lemma : 
{\footnotesize \begin{equation}\label{gronwall}
\left|\left|\left( \begin{array}{c}
u_{1}(x)-u_{2}(x) \\
u_{1}'(x)-u_{2}'(x)
\end{array}\right)\right|\right| \leq \left( \int_{x}^{y} ||V_{1}(t)-V_{2}(t)||\ ||u_{1}(t)||dt \right) \exp \left( \int_{x}^{y} (||V_{2}(t)||+1) dt\right)
\end{equation}}

But by lemma \ref{lemmaapriori1}, for all $t\in[y,x]$ : 
{\small $$||u_{1}(t)||^{2} \leq ||u_{1}(t)||^{2}+||u_{1}'(t)||^{2} \leq \left( ||u_{1}(y)||^{2}+||u_{1}'(y)||^{2}\right) \exp \left( \int_{x}^{y} (||V_{1}(s)||+1) ds\right)$$}
So : 
$$||u_{1}(t)|| \leq \left( ||u_{1}(y)||^{2}+||u_{1}'(y)||^{2}\right)^{\frac{1}{2}} \exp \left(\frac{1}{2} \int_{x}^{y} (||V_{1}(s)||+1) ds\right)$$
We put this in (\ref{gronwall}) : 
$$\left( ||u_{1}(x)-u_{2}(x)||^{2}+||u_{1}'(x)-u_{2}'(x)||^{2} \right)^{\frac{1}{2}} \qquad \qquad \qquad \qquad \qquad \qquad \qquad\qquad $$
$$\leq \left(  ||u_{1}(y)||^{2}+||u_{1}'(y)||^{2} \right)^{\frac{1}{2}} \exp \left( \int_{\min(x,y)}^{\max(x,y)} \frac{1}{2}||V_{1}(t)||+\frac{1}{2} +||V_{2}(t)||+1 dt\right)\times$$ 
$$\qquad \qquad \qquad \qquad \qquad \qquad \qquad \qquad \qquad \qquad\int_{\min(x,y)}^{\max(x,y)} ||V_{1}(t)-V_{2}(t)||dt$$
And we have finished the proof because : $\frac{1}{2}||V_{1}(t)||+\frac{1}{2}\leq ||V_{1}(t)||+1$.
\end{proof}

\noindent \emph{Notation: }Let $u^{1},\ldots,u^{2N}$ be solutions of (\ref{system2nd}) with initial conditions:
\begin{equation}\label{system2CI}
\left(\begin{array}{c}
u^{1}(n,E) \\
(u^{1})'(n,E)
\end{array}\right)
= \left(\begin{array}{c} 
1 \\
0 \\
\vdots \\
0
\end{array}\right), \ldots, \left(\begin{array}{c}
u^{2N}(n,E) \\
(u^{2N})'(n,E)
\end{array}\right) = \left( \begin{array}{c} 
0 \\
\vdots \\
0 \\
1
\end{array}\right)
\end{equation}
Then the transfer matrix $A_{n}^{\omega}(E)$ has the expression:
\begin{equation}\label{tmatexp}
A_{n}^{\omega}(E)=\left(\begin{array}{ccc}
u^{1}(n+1,E) & \ldots & u^{2N}(n+1,E)\\
(u^{1})'(n+1,E) & \ldots & (u^{2N})'(n+1,E)
\end{array}\right)
\end{equation}

\begin{proof}[of theorem \ref{holderlyap}]
We start by proving (\ref{estim1}). Let ${}^t(u^{i}(n+1,E)\ (u^{i}){'}(n+1,E))$ be the column of $A_{n}^{\omega}(E)$ of maximal norm. Then:
$$||A_{n}^{\omega}(E)||^{2}=||u^{i}(n+1,E)||^{2}+||(u^{i})'(n+1,E)||^{2}$$
Applying lemma \ref{lemmaapriori1} with $x=n+1$ and $y=n$ one gets:
$$||u^{i}(n+1,E)||^{2}+||(u^{i})'(n+1,E)||^{2}\qquad \qquad \qquad \qquad \qquad \qquad \qquad \qquad \qquad \qquad $$
$$ \leq \left( ||u^{i}(n,E)||^{2}+||(u^{i})'(n,E)||^{2} \right) \exp \left( \int_{n}^{n+1} ||V_{\omega}(t)-E||+1\dd t\right)$$
But due to (\ref{system2CI}) we have: $||u^{i}(n,E)||^{2}+||(u^{i})'(n,E)||^{2}=1$. We also have that $x\mapsto V_{\omega}(x)$ is $1$-periodic. Thus:
$$\int_{n}^{n+1} ||V_{\omega}(t)-E||+1\dd t = \int_{0}^{1} ||V_{\omega}(t)-E||+1\dd t \leq\left( \sup_{t\in [0,1]} ||V_{\omega}(t)||\right)+ |E| +1 $$
But $V_{\omega}$ being uniformly bounded on $x$ and $\omega$, there exists $C_{1}>0$ independent of $\omega,n$ and $E$ such that: 
$$\left( \sup_{t\in [0,1]} ||V_{\omega}(t)||\right) \leq C_{1}$$
Then: $$||A_{n}^{\omega}(E)||^{2} \leq \exp( C_{1} + |E| +1)$$
As $I$ is compact, $|E|$ is also bounded and so there exists $\tilde{C}_{2}>0$ independent of $\omega,n$ and $E$ such that: $\exp( C_{1} + |E| +1) \leq \tilde{C}_{2}$. Finally, we use that for every $p\in \{ 1,\ldots,2N \}$ and for every $M\in \mathrm{GL}_{2N}(\R)$: $||\wedge^{p} M||\leq ||M||^{p}$. Applying it to $M=A_{n}^{\omega}(E)$, we obtain (\ref{estim1}).
\vskip 2mm

To prove (\ref{estim2}) we first prove it for $p=1$. Let $E,E'\in I$. First there exists $i\in \{ 1,\ldots,2N\}$ such that:
$$||A_{n}^{\omega}(E)-A_{n}^{\omega}(E')||=\left|\left|\left(\begin{array}{c}
u^{i}(n+1,E) \\
(u^{i})'(n+1,E)
\end{array}\right)-\left(\begin{array}{c}
u^{i}(n+1,E') \\
(u^{i})'(n+1,E')
\end{array}\right)\right|\right|\qquad \qquad$$
$$\qquad \qquad \qquad\leq \left|\left|\left(\begin{array}{c}
u^{i}(n,E) \\
(u^{i})'(n,E)
\end{array}\right)\right|\right|\left( \int_{n}^{n+1} ||V_{\omega}(t)-E-(V_{\omega}(t)-E')||\dd t \right)\times $$
$$\qquad \qquad \qquad \qquad \exp \left( \int_{n}^{n+1} ||V_{\omega}(t)-E||+||(V_{\omega}(t)-E')||+2\ \dd t \right)$$

by lemma \ref{lemmaapriori2}. Thus:
\begin{eqnarray}
||A_{n}^{\omega}(E)-A_{n}^{\omega}(E')|| &\leq & |E-E'|\exp\left( \int_{0}^{1} 2||V_{\omega}(t)||+|E|+|E'|+2 \dd t \right) \nonumber \\
 & \leq & |E-E'| \exp(2C_{1}+2+2\ell(I)) \nonumber \\
 & \leq & \tilde{C}_{3} |E-E'| \nonumber 
\end{eqnarray}
with $\tilde{C}_{3}$ independent of $n,\omega$ and $E$. Now for $p\geq 1$ we use the following estimate valid for $M,N\in \mathrm{GL}_{2N}(\R)$ and $p\in \{ 1,\ldots,2N \}$:
$$||\wedge^{p}M-\wedge^{p}N|| \leq ||N-M||(||N||^{p-1}+||M||.||N||^{p-2}+\ldots +||M||^{p-1})$$
It is a direct computation (see \cite{boumazathese} p.$118$ for details). Applying it to $M=A_{n}^{\omega}(E)$ and $N=A_{n}^{\omega}(E')$ one gets:
$$||\wedge^{p}A_{n}^{\omega}(E)-\wedge^{p}A_{n}^{\omega}(E')|| \leq p C_{2}^{p-1} \tilde{C}_{3} |E-E'|$$
and $C_{3}=p C_{2}^{p-1} \tilde{C}_{3}$ is independent of $n,\omega,E$ and $E'$.
\vskip 2mm

We have checked $(ii)$ and $(iii)$ in theorem \ref{holderlyapgen} and $(i)$ is an assumption in theorem \ref{holderlyap}. Therefore we can apply theorem \ref{holderlyapgen} to have the H\"older continuity on $I$ of the Lyapounov exponents associated to $H_{A}(\omega)$. 
\end{proof}

\section{H\"older continuity of the IDS}\label{section4}

\subsection{Kotani's $w$ function}

We start by introducing the $w$ function of Kotani as defined in \cite{kotanis} for matrix-valued Schr\"odinger operators. For this, we first have to define the $m$-functions associated to such operators. We follow \cite{kotanis} and we will refer to this article for all proofs of this paragraph. Let $\C_{+}$ denote the half upper plane $\{z\in \C \ |\ \mathrm{Im}(z)>0 \}$ and $\C_{-}$ the lower half plane $\{z\in \C \ |\ \mathrm{Im}(z)<0 \}$.

\begin{prop}\label{Fplus}
Let $E\in \C_{+}\cup \C_{-}$. We fix $\omega \in \Omega$. Then there exists a unique function  $x\mapsto F_{+}(x,E)$ with values in $\mathcal{M}_{\mathrm{N}}(\C)$ (respectively $x\mapsto F_{-}(x,E)$) satisfying : 
$$-F_{+}''+V_{\omega}F_{+}=EF_{+},\ F_{+}(0,E)=I,\ \mathrm{and}\ \int_{0}^{\infty} ||F_{+}(x,E)||^{2}dx <+\infty$$
respectively : 
$$-F_{-}''+V_{\omega}F_{-}=EF_{-},\ F_{-}(0,E)=I,\ \mathrm{and}\ \int_{-\infty}^{0} ||F_{-}(x,E)||^{2}dx <+\infty$$
\end{prop}

\begin{proof}
See \cite{kotanis}, Corollary $2.2$.
\end{proof}

\begin{Def}
For $E\in \C_{+} \cup \C_{-}$ we define the $m$-functions $M_{+}$ and $M_{-}$ associated to $H_{A}(\omega)$ by : 
$$M_{+}(E)=\frac{d}{dx} F_{+}(x,E)|_{x=0} \ \mathrm{and}\ M_{-}(E)=-\frac{d}{dx} F_{-}(x,E)|_{x=0}$$
\end{Def}

With these functions we can give the expression of the Green kernel of the resolvant of $H_{A}(\omega)$.

\begin{prop}\label{greenker}
Let $E\in \C_{+} \cup \C_{-}$. Then  $(H_{A}(\omega)-E)^{-1}$ has a continuous integral kernel $G_{E}(x,y,\omega)$ given by : 
$$G_{E}(x,y,\omega) = \left\lbrace \begin{array}{lcc}
-F_{-}(x)(M_{+}+M_{-})^{-1}\ ^{t}F_{+}(y) & \mathrm{if} & x\leq y \\
-F_{+}(x)(M_{+}+M_{-})^{-1}\ ^{t}F_{-}(y) & \mathrm{if} & y\leq x 
\end{array} \right.$$
\end{prop}

\begin{proof}
See \cite{kotanis}, Theorem $3.2$.
\end{proof}

We can now define the $w$ function of Kotani. This function will be the link between the Lyapounov exponents and the IDS. Indeed, its real part will be the sum of the $N$ positive Lyapounov exponents while its imaginary part will tend to $\pi N(E)$ when $E$ tends to the real line.

\begin{Def}
Let $E\in \C_{+} \cup \C_{-}$. We define the $w$ function of Kotani by: 
$$w(E) = \frac{1}{2} \E (\mathrm{Tr}(M_{+}(E)+M_{-}(E)))$$
\end{Def}

Then the $w$ function has the following properties:

\begin{prop}\label{wprop}
For $E \in \C_{+} \cup \C_{-}$:
\begin{enumerate}[\rm(i)]
\item
$w(E)=\mathbb{E}(\mathrm{Tr}(M_{+}(E)))=\mathbb{E}(\mathrm{Tr}(M_{-}(E)))$
\item
$\frac{\dd}{\dd E}w(E)=\mathbb{E}(\mathrm{Tr}(G_{E}(0,0,\omega)))$
\item
$-\mathrm{Re}\ w(E) = (\gamma_{1}+\ldots +\gamma_{N})(E)$ 
\item
$\E \left( \mathrm{Tr}(\mathrm{Im}\ M_{\pm}(E,\omega)^{-1})\right)= -\frac{2\mathrm{Re}\ w(E)}{\mathrm{Im} E}=\frac{2(\gamma_{1}+\ldots +\gamma_{N})(E)}{\mathrm{Im} E}$
\end{enumerate}
\end{prop}

\begin{proof}
See \cite{kotanis}, Theorem $6.2C$.
\end{proof}

In point $(iii)$ we have to precise that the formula:
$$(\gamma_{1}(E)+\ldots +\gamma_{N})(E))= \lim_{n \to \infty} \frac{1}{n}
\mathbb{E}(\log ||\wedge^{N} (A_{n-1}^{\omega}\ldots A_{0}^{\omega})
||)$$
makes sense for every $E\in \C$.
\vskip 2mm

We can now generalize results of harmonic analysis of the $w$ function presented in the case of scalar-valued Schr\"odinger operators by Kotani in \cite{kotani83} to the case of matrix-valued Schr\"odinger operators.

First we introduce the space of Herglotz functions:
$$\mathcal{H}=\{ h \ |\ h\ \mathrm{is}\ \mathrm{holomorphic}\ \mathrm{on}\ \C_{+} \ \mathrm{and}\ h\ :\ \C_{+}\to \C_{+} \}$$
Then we define a subspace of $\mathcal{H}$:
$$\mathcal{W}= \{ w\in \mathcal{H}\ |\ w,\ w',\ -\mathrm{i}w\in \mathcal{H} \} $$

\begin{prop}
The Kotani's function $w$ is in $\mathcal{W}$.
\end{prop}

\begin{proof}
First, as $H_{A}(\omega)$ is self-adjoint, its spectrum is included in $\R$ and $E\mapsto M_{+}(E)$ is holomorphic on $\C \setminus \R$ and so is $E\mapsto \mathrm{Tr}(M_{+}(E))$. If $\mathrm{Im}E>0$, by Proposition $2.3$ (a) in \cite{kotanis}, one has:
$$\mathrm{Im}\ M_{+}(E) = (\mathrm{Im}E)\int_{0}^{+\infty} F_{+}(x,E)^{*}F_{+}(x,E) >0$$
Thus, $E\mapsto \mathrm{Tr}(M_{+}(E))$ is in $\mathcal{H}$ and $w\in \mathcal{H}$.

Then by proposition \ref{wprop} $(ii)$, $w'(E)=\mathbb{E} (\mathrm{Tr}(G_{E} (0,0,\omega)))$. But $G_{E}(0,0,\omega)$ is holomorphic away from the spectrum of $H_{A}(\omega)$ and so is $\mathrm{Tr}(G_{E}(0,0,\omega))$. If $\mathrm{Im}E>0$, then the operator $\mathrm{Im}(H_{A}(\omega)-E)^{-1}$ is a positive definite operator and $\mathrm{Im}\mathrm{Tr}(G_{E}(0,0,\omega))>0$. Then $\mathrm{Im} w'(E)= \mathrm{Im}\mathrm{Tr}(G_{E}(0,0,\omega))>0$ and $w'\in \mathcal{H}$.

Finally, $-\mathrm{i}w$ is holomorphic on $\C_{+}$ as $w$ is. If $E\in \C_{+}$: 
$$\mathrm{Im}(-\mathrm{i}w(E))=- \mathrm{Re}\ w(E)=(\mathrm{Im}E) \E(\mathrm{Tr}(\mathrm{Im}\ M_{+}(E,\omega)^{-1})) $$
by proposition \ref{wprop} $(iv)$. But if $E\in \C_{+}$, $\mathrm{Tr}(\mathrm{Im}\ M_{+}(E,\omega)^{-1})>0$ and then $\mathrm{Im}(-\mathrm{i}w(E))>0$. Therefore, $-\mathrm{i}w\in \mathcal{H}$.
\end{proof}

\subsection{A Thouless formula}

Let $\mathfrak{n}$ be the measure defined in proposition \ref{laplaceIDS}.

\begin{prop}\label{idstrace}
\begin{equation}\label{idstraceq}
\forall E\in \C \setminus \R,\ \E(\mathrm{Tr}\ G_{E}(0,0,\omega)) = \int_{\R} \frac{\dd\mathfrak{n}(E')}{E'-E}
\end{equation}
\end{prop}

\begin{proof}
As $\R$ is a limit of bounded Borel sets and the Dirac distribution at $0$, $\delta_{0}$, can be approached by compactly supported continuous functions, positives and of $L^{2}$-norm equal to $1$, using proposition \ref{idsEtrace} we have:
$$\int_{\R} \frac{\mathrm{d}\mathfrak{n}(E')}{E'-E} = \int_{\R} \frac{1}{E'-E}\dd \E \left( \mathrm{Tr}(<\delta_{0},E_{H_{A}(\omega)}((-\infty,E'])\delta_{0}>) \right)$$
Then applying the spectral theorem to the self-adjoint operator $H_{A}(\omega)$:
\begin{eqnarray}
\int_{\R} \frac{\dd \mathfrak{n}(E')}{E'-E} & = & \E\left( \mathrm{Tr}(\int_{\R} \frac{1}{E'-E} \dd<\delta_{0},E_{H_{A}(\omega)}((-\infty,E'])\delta_{0}>) \right) \nonumber \\
 & = & \E\left( \mathrm{Tr}(<\delta_{0},\left(\int_{\R} \frac{1}{E'-E} \dd E_{H_{A}(\omega)}((-\infty,E'])\right)\delta_{0}>) \right) \nonumber \\
 & = & \E\left( \mathrm{Tr}(<\delta_{0},(H_{A}(\omega)-E)^{-1}\delta_{0}>) \right) \nonumber \\
 & = & \E\left( \mathrm{Tr}(G_{E}(0,0,\omega)) \right) \nonumber
\end{eqnarray}
\end{proof}

With this proposition, we can express the imaginary part of $w$ in terms of the IDS $N(E)$.

\begin{prop}\label{imw}
\begin{equation}\label{imweq}
\forall E\in \R,\ \lim_{a \to 0^{+}} \mathrm{Im}\ w(E+\mathrm{i}a)=\pi N(E)
\end{equation}
\end{prop}

\begin{proof}
First, by proposition \ref{wprop} $(ii)$:
$$\forall z\in \C \setminus \R,\ w'(z)=\E(\mathrm{Tr}(G_{z}(0,0,\omega)))$$
Then, we can apply proposition \ref{idstrace}:
\begin{eqnarray}
\forall z\in \C \setminus \R,\ w'(z) & = & \int_{\R} \frac{\mathrm{d}\mathfrak{n}(E')}{E'-z} \nonumber \\
 & = & \int_{\R} \frac{N(E')}{(E'-z)^{2}}~\mathrm{d}E' \nonumber
\end{eqnarray}
by integrating by parts. Then by integrating this expression, there exists a constant $c\in \C$ such that:
\begin{equation}\label{integralew}
w(z)=c+\int_{\R} \frac{1+E'z}{(E'-z)(1+E'^{2})} N(E')\mathrm{d}E'
\end{equation}
But if $z\in \R$ is not in the spectrum of $H_{A}(\omega)$ then $w(z)\in \R$ (see \cite{carmonastflour}, lemma $5.10$, p$84$). Thus we must have $c\in \R$. Then, taking imaginary part in (\ref{integralew}) and writing for $z\in \C_{+}$, $z=E+\mathrm{i}a$, $E\in \R$, $a>0$:
\begin{eqnarray}
\mathrm{Im}\ w(E+\mathrm{i}a) & = & a \int_{\R} \frac{N(E')}{(E'-E)^{2}+a^{2}}~\mathrm{d}E' \nonumber \\
 & = & \int_{\R} \frac{N(E+au)}{1+u^{2}}~\mathrm{d}u \nonumber
\end{eqnarray}
where $u=\frac{E'-E}{a}$. But $N(E)$ being a repartition function, it is right continuous and so:
$$\forall E\in \R,\ \lim_{a\to 0^{+}}\mathrm{Im}\ w(E+\mathrm{i}a)= N(E) \int_{\R} \frac{1}{1+u^{2}}~ \mathrm{d}u = \pi N(E)$$
\end{proof}

We have an analoguous proposition for the real part of $w(E)$.

\begin{prop}\label{rew}
For Lebesgue-almost every $E$ in $\R$, we have: 
\begin{equation}\label{reweq}
\lim_{a \to 0^{+}} \mathrm{Re}\ w(E+\mathrm{i}a)=-(\gamma_{1}+\ldots +\gamma_{N})(E)
\end{equation}
Moreover, if $I\subset \R$ is an interval on which $E\mapsto -(\gamma_{1}+\ldots +\gamma_{N})(E)$ is continuous then (\ref{reweq}) holds for every $E\in I$.
\end{prop}

\begin{proof}
First by proposition \ref{wprop} $(iii)$, we have:
\begin{equation}\label{reweq1}
\forall z\in \C \setminus \R,\ \mathrm{Re}\ w(z)=-(\gamma_{1}+\ldots +\gamma_{N})(z)
\end{equation}
The function $z\mapsto -(\gamma_{1}+\ldots +\gamma_{N})(z)$ is subharmonic (see \cite{cs83}) and so for almost every $E$ in $\R$ the following limit exists:
\begin{equation}\label{reweq2}
\lim_{a \to 0} (\gamma_{1}+\ldots +\gamma_{N})(E+\mathrm{i}a) = (\gamma_{1}+\ldots +\gamma_{N})(E)
\end{equation}
Let $E$ be a real number such that (\ref{reweq2}) holds. Then setting $z=E+\mathrm{i}a$ with $a>0$ in (\ref{reweq1}) one gets the existence of the following limit:
\begin{equation}\label{reweq3}
\lim_{a \to 0^{+}} \mathrm{Re}\ w(E+\mathrm{i}a)=-(\gamma_{1}+\ldots +\gamma_{N})(E)
\end{equation}
Moreover, if $I$ is an interval on which $E\mapsto (\gamma_{1}+\ldots +\gamma_{N})(E)$ is continuous, the relation (\ref{reweq3}) holds for every $E$ in $I$ as it holds for almost every $E\in I$.
\end{proof}
\vskip 2mm

Now we can prove a Thouless formula adapted to matrix-valued continuous Schr\"odinger operators. As $(\gamma_{1}+\ldots +\gamma_{N})(E)$ and $N(E)$ are respectively the real and imaginary part of the function $w$ which lies in $\mathcal{W}$, the harmonic analysis developed in \cite{kotani83} says these two functions are linked by an integral relation.

\begin{thm}[\textbf{Thouless formula}]\label{thouless}
For almost every $E\in \R$ we have : 
\begin{equation}\label{thoulesseq}
(\gamma_{1}+\ldots+\gamma_{N})(E)= -\alpha+\int_{\R} \log \left( \left| \frac{E'-E}{E'-\mathrm{i}}\right| \right)~\dd \mathfrak{n}(E')
\end{equation}
where $\alpha$ is a real number independent of $E$ and $\mathfrak{n}$ is the measure of which the IDS $E\mapsto N(E)$ is the repartition function.
Moreover, if  $I\subset \R$ is an interval on which $E\mapsto -(\gamma_{1}+\ldots +\gamma_{N})(E)$ is continuous then (\ref{thoulesseq}) holds for every $E\in I$.
\end{thm}

\begin{proof}
As $w\in \mathcal{W}$, we can apply to $w$ the lemma $7.7$ in \cite{kotani83}. In particular, using also proposition \ref{imw}, we have:
\begin{equation}\label{repintw}
\forall z\in \C \setminus \R,\ w(z)=w(\mathrm{i})+\int_{\R} \log \left( \frac{E'-\mathrm{i}}{E'-z} \right)~\dd\mathfrak{n}(E')
\end{equation}
Then:
\begin{equation}\label{repintwre}
\mathrm{Re}\ w(z) = \mathrm{Re}\ w(\mathrm{i}) + \int_{\R} \log \left( \left| \frac{E'-\mathrm{i}}{E'-z}\right| \right)~\mathrm{d}\mathfrak{n}(E')
\end{equation}
Let $z=E+\mathrm{i}a$ with $E\in \R$ such that (\ref{reweq}) holds and $a>0$. Then when $a$ goes to $0$, by proposition \ref{rew} we have:
\begin{equation}
-(\gamma_{1}+\ldots+\gamma_{N})(E)=\mathrm{Re}\ w(\mathrm{i}) + \int_{\R} \log \left( \left| \frac{E'-\mathrm{i}}{E'-E}\right| \right)~\mathrm{d}\mathfrak{n}(E')
\end{equation}
If we set $\alpha=\mathrm{Re}\ w(\mathrm{i})$ we finally get (\ref{thoulesseq}) for every $E$ in $\R$ such that (\ref{reweq}) holds, \emph{i.e} for almost every $E$ in $\R$. Then if $I$ is an interval on which $E\mapsto (\gamma_{1}+\ldots +\gamma_{N})(E)$ is continuous, by proposition \ref{rew}, (\ref{thoulesseq}) will hold for every $E$ in $I$.
\end{proof}
\vskip 3mm

We can now use this Thouless formula to prove that the IDS $E\mapsto N(E)$ has the same regularity as the Lyapounov exponents.

\subsection{Local H\"older continuity of the IDS}

We start by a quick review of the Hilbert transform and its main properties. For the proofs we refer to \cite{neri}, chapter $3$.

\begin{Def}
If $\psi\in L^{2}(\R)$, its Hilbert transform is the function defined on $\R$ by: 
$$(T\psi)(x)= \lim_{\varepsilon \to 0^{+}} \frac{1}{\pi} \int_{|x-t|>\varepsilon} \frac{\psi(t)}{x-t}~\dd t$$
\end{Def}

\begin{prop}\label{hilbertprop}
Let $\psi\in L^{2}(\R)$.
\begin{enumerate}[\rm(i)]
\item Then $T^{2}\psi(x)=-\psi(x)$ for almost every $x$ in $\R$.
\item If $\psi$ is H\"older continuous on the interval $[x_{0}-a,x_{0}+a]$, $a>0$, then $T\psi$ is H\"older continuous on the interval $[x_{0}-\frac{a}{2},x_{0}+\frac{a}{2}]$.
\end{enumerate}
\end{prop}

Now we can prove the following result of regularity of the IDS.

\begin{thm}\label{holderids}
Let $I$ be a compact interval in $\R$ and $\tilde{I}$ be an open interval, $I\subset \tilde{I}$. We assume that the potential $V_{\omega}$ in $H_{A}(\omega)$ for $d=1$ and $N\geq 1$ is such that the group $G_{\mu_{E}}$ associated to the transfer matrices of $H_{A}(\omega)$ is $p$-contracting and $L_p$-strongly irreducible for every $p\in \{1,\ldots ,N\}$ and every $E\in \tilde{I}$. Then the IDS associated to $H_{A}(\omega)$ is H\"older continuous on $I$.
\end{thm}

\begin{proof}
First, the application $E'\mapsto \log \left( \left| \frac{E'-E}{E'-\mathrm{i}} \right| \right)$ is $\mathfrak{n}$-integrable on $\R$. Indeed the renormalisation term $E'-\mathrm{i}$ at the denominator balances the fact that the support of $\mathfrak{n}$ is non-compact. Thus, we have:
\begin{equation}
\forall E\in \R,\ \lim_{\varepsilon \to 0^{+}} \int_{E-\varepsilon}^{E+\varepsilon} \left|\log \left( \left| \frac{E'-E}{E'-\mathrm{i}} \right| \right)  \right|  ~\dd\mathfrak{n}(E')=0
\end{equation}
from which we deduce that:
\begin{equation}\label{idscont}
\forall E\in \R,\ \lim_{\varepsilon \to 0^{+}} |\log(\varepsilon)| (N(E+\varepsilon)-N(E-\varepsilon))=0 
\end{equation}
It implies that $E\mapsto N(E)$ is continuous on $\R$. Let $E_{0}\in I$ be fixed and $a>0$ such that $[E_{0}-4a,E_{0}+4a] \subset \tilde{I}$. Then, by theorem \ref{thouless}, for $E\in ]E_{0}-4a,E_{0}+4a[$:
{\scriptsize $$(\gamma_{1}+\ldots +\gamma_{N})(E)+\alpha - \int_{|E'-E_{0}|>4a} \log \left( \left| \frac{E'-E}{E'-\mathrm{i}} \right| \right) ~\mathrm{d}\mathfrak{n}(E') = \int_{E_{0}-4a}^{E_{0}+4a} \log \left( \left| \frac{E'-E}{E'-\mathrm{i}} \right| \right) ~\mathrm{d}\mathfrak{n}(E')$$}
Then:
 $$\int_{E_{0}-4a}^{E_{0}+4a} \log \left( \left| \frac{E'-E}{E'-\mathrm{i}} \right| \right) ~\mathrm{d} \mathfrak{n}(E') = \lim_{\varepsilon \to 0^{+}} \left( \int_{E_{0}-4a}^{E-\varepsilon} \log|E'-E| ~\mathrm{d}\mathfrak{n}(E') \right. \qquad \qquad$$
 $$\left.\qquad \qquad \qquad + \int_{E+\varepsilon}^{E_{0}+4a} \log|E'-E| ~\mathrm{d}\mathfrak{n}(E') \right)  - \frac{1}{2}\int_{E_{0}-4a}^{E_{0}+4a} \log (1+(E')^{2}) ~\mathrm{d}\mathfrak{n}(E')  $$
We set:
$$\mathcal{I}(E_{0})=\frac{1}{2}\int_{E_{0}-4a}^{E_{0}+4a} \log (1+(E')^{2}) ~\mathrm{d}\mathfrak{n}(E')$$ 
Then, integrating by parts the first two integrals leads to:
$$\int_{E_{0}-4a}^{E_{0}+4a} \log \left( \left| \frac{E'-E}{E'-\mathrm{i}} \right| \right) ~\mathrm{d}\mathfrak{n}(E')  \qquad \qquad \qquad \qquad \qquad  \qquad \qquad \qquad \qquad $$
{\small \begin{eqnarray}
 & = & \lim_{\varepsilon \to 0^{+}} \left[ [N(E')\log|E'-E|]_{E_{0}-4a}^{E-\varepsilon} - \int_{E_{0}-4a}^{E-\varepsilon} \frac{N(E')}{E-E'} dE' +  [N(E')\log|E'-E|]_{E+\varepsilon}^{E_{0}+4a} \right. \nonumber \\
 & & \left. - \int_{E+\varepsilon}^{E_{0}+4a} \frac{N(E')}{E'-E} dE' \right] -\mathcal{I}(E_{0}) \nonumber
\end{eqnarray} }
We set $\psi(E)=N(E)\chi_{\{|E-E_{0}| \leq 4a \}}\in L^{2}(\R)$. By definition of the Hilbert transform:
$$\int_{E_{0}-4a}^{E_{0}+4a} \log|E'-E| ~\mathrm{d}\mathfrak{n}(E') \qquad \qquad \qquad \qquad \qquad \qquad \qquad  \qquad \qquad \qquad \qquad  $$
{\small \begin{eqnarray}
& = & \pi (T\psi)(E) +  \lim_{\varepsilon \to 0^{+}} \left[ (N(E-\varepsilon)-N(E+\varepsilon))\log \varepsilon +N(E_{0}+4a)\log|E_{0}-E+4a| \right. \nonumber \\
 & & \left. \qquad \qquad \qquad \qquad  \qquad \qquad \qquad \qquad - N(E_{0}-4a)\log|E_{0}-E-4a| \right] -\mathcal{I}(E_{0}) \nonumber \\
 & = & \pi (T\psi)(E)+N(E_{0}+4a)\log|E_{0}-E+4a| - N(E_{0}-4a)\log|E_{0}-E-4a| -\mathcal{I}(E_{0}) \nonumber
\end{eqnarray} }
by (\ref{idscont}). We finally get:
{\small \begin{eqnarray}
\pi (T\psi)(E) & = & (\gamma_{1}+\ldots +\gamma_{N})(E)+\alpha - \int_{|E'-E_{0}|>4a} \log \left( \left| \frac{E'-E}{E'-\mathrm{i}} \right| \right)~\mathrm{d}\mathfrak{n}(E')-  \nonumber \\
& & N(E_{0}+4a)\log|E_{0}-E+4a|+ N(E_{0}-4a)\log|E_{0}-E-4a| +\mathcal{I}(E_{0})\nonumber \\
& = & (\gamma_{1}+\ldots +\gamma_{N})(E)+\alpha - \int_{|E'-E_{0}|\geq 4a} \log \left( \left| \frac{E'-E}{E'-\mathrm{i}} \right| \right)~\mathrm{d}\mathfrak{n}(E')+\mathcal{I}(E_{0}) \nonumber
\end{eqnarray} }
But as $[E_{0}-4a,E_{0}+4a]\subset I \subset \tilde{I}$, $E\mapsto (\gamma_{1}+\ldots +\gamma_{N})(E)$ is H\"older continuous on $[E_{0}-4a,E_{0}+4a]$ by theorem \ref{holderlyap}. Moreover, $E\mapsto \int_{|E'-E_{0}|\geq 4a} \log \left( \left| \frac{E'-E}{E'-\mathrm{i}} \right| \right)~\mathrm{d}\mathfrak{n}(E')$ is H\"older continuous of order $1$ on the interval $]E_{0}-4a,E_{0}+4a[$.

Then $T\psi$ is H\"older continuous on every compact interval included in $]E_{0}-4a,E_{0}+4a[$, in particular it is H\"older continuous on $[E_{0}-2a,E_{0}+2a]$. Thus by proposition \ref{hilbertprop} $(ii)$, $T^{2}\psi$ is H\"older continuous on $[E_{0}-a,E_{0}+a]$. But by proposition \ref{hilbertprop} $(i)$ and by continuity of $E\mapsto N(E)$ (by (\ref{idscont})), we have:
$$\forall E\in [E_{0}-a,E_{0}+a],\ (T^{2}\psi)(E)=-N(E)$$ 
Then $E\mapsto N(E)$ is H\"older continuous on $[E_{0}-a,E_{0}+a]$. But $I$ being compact, it can be covered by a finite number of intervals $]E_{0}-a,E_{0}+a[\subset \tilde{I}$ with $E_{0}\in I$. Thus, $E\mapsto N(E)$ is H\"older continuous on $I$.
\end{proof}
\vskip 3mm

The H\"older continuity of the Lyapounov exponents and of the IDS relies on the assumptions of $p$-contractivity and $L_p$-strong irreducibility for every $p\in \{ 1,\ldots, N\}$ made on $G_{\mu_{E}}$. But, for arbitrary potential $V_{\omega}$, we do not know if these assumptions are verified or not. In the next section we will present a first example of continuous matrix-valued Anderson model for which these assumptions are verified.

\section{Anderson model on two coupled strings}\label{section5}

We will now see how to apply theorem \ref{holderids} to a particular case of $H_{A}(\omega)$, which is the following operator:
\begin{equation}
H_{AB}(\omega)=-\frac{\dd^{2}}{\dd x^{2}}\otimes I_{2} +\left( \begin{array}{cc}
0 & 1 \\ 
1 & 0
\end{array}\right)+\sum_{n\in \Z} \left( \begin{array}{cc}
\omega_{1}^{(n)} \chi_{[0,1]}(x-n) & 0 \\
0 & \omega_{2}^{(n)} \chi_{[0,1]}(x-n)
\end{array}\right) \nonumber
\end{equation}

Here, $\chi_{[0,1]}$ denotes the characteristic function of the interval $[0,1]$ and $(\omega_{1}^{(n)})_{n\in \Z}$ and $(\omega_{2}^{(n)})_{n\in \Z}$ are two independent sequences of \emph{i.i.d.} random variables with common law $\nu$ such that $\{ 0,1\} \subset \supp \nu$. This operator is a bounded perturbation of $(-\frac{\dd^{2}}{\dd x^{2}})
\otimes I_{2}$ and thus self-adjoint on the Sobolev space $H^{2}(\R)\otimes \C^{2}$.
\vskip 2mm

For the operator $H_{AB}(\omega)$, we have the following result:

\begin{thm}\label{idshab}
The Integrated Density of States $N(E)$ associated to $H_{AB}(\omega)$ exists for every $E\in \R$. Moreover, there exists a discrete subset $\mathcal{S}_{B} \subset \R$ such that for every compact interval $I\subset (2,+\infty)\setminus \mathcal{S}_{B}$, the function $E\mapsto N(E)$ is H\"older continuous on $I$.
\end{thm}

According to theorem \ref{holderids}, we only have to prove that there exists a discrete subset $\mathcal{S}_{B} \subset \R$ such that for every $E\in (2,+\infty)\setminus \mathcal{S}_{B}$, the group $G_{\mu_{E}}$ associated to the transfer matrices of $H_{AB}(\omega)$ is $p$-contracting and $L_p$-strongly irreducible for $p\in \{1,2\}$. It has already been proved in a previous article of the author, \cite{boumaza}, and we will only give here the outlines of the proof and some comments.

To prove that an explicit  group is $p$-contracting and $L_p$-strongly irreducible can be very complicated. It has been done in \cite{DSS02} for the case of a scalar-valued continuous Anderson model, but their proof relies on properties of reflection and transmission coefficients which no longer holds in the matrix-valued case. In the case of a discrete matrix-valued  Anderson model, a more algebraic approach has been successfully used by Gol'dsheid and Margulis in \cite{GM89}. We follow here this approach and adapt it to the case of continuous matrix-valued Anderson models. It is based on the following criterion:

\begin{thm}[Gol'dsheid, Margulis \cite{GM89}]\label{zariski}
If a subgroup $G$ of $\mathrm{Sp}_{N}(\R)$ is dense for the Zariski topology in $\mathrm{Sp}_{N}(\R)$ then it is $p$-contracting and $L_p$-strongly irreducible for every $p\in \{ 1,\ldots, N\}$. 
\end{thm}

In the case of a discrete matrix-valued Anderson model, the transfer matrices have a simple enough expression to make possible a direct construction of the Zariski closure of the group $G_{\mu_{E}}$ generated by these transfer matrices. And so it can be proved that for every $E\in \R$, $G_{\mu_{E}}$ is Zariski dense in $\mathrm{Sp}_{N}(\R)$.
\vskip 2mm

In our case, the transfer matrices associated to $H_{AB}(\omega)$, even if they are still explicit, are complicated enough to not allow a direct reconstruction of the Zariski closure of $G_{\mu{E}}$ for every $E$ except those in a discrete set. It is due to the fact that $E$ and the $\omega_{i}$'s are not separated in the expressions of these transfer matrices. A direct reconstruction of the Zariski closure of $G_{\mu_{E}}$ is in fact possible, but only for values of $E$ away from a dense countable subset of $\R$, as shown in \cite{stolzboumaza}. It leads to the impossibility to find an interval of values of $E$ such that $G_{\mu_{E}}$ is $p$-contracting and $L_p$-strongly irreducible and makes it impossible to apply theorem \ref{holderids}.

The idea in \cite{boumaza}, to improve the result of \cite{stolzboumaza}, is to combine the criterion of Gol'dsheid and Margulis to a recent result of Breuillard and Gelander on Lie groups:

\begin{thm}[Breuillard, Gelander \cite{breuillarda}]\label{breuillard}
Let $G$ be a real, connected, semisimple Lie group, whose Lie algebra is $\mathfrak{g}$. 

Then there exists a neighborhood $\mathcal{O}$ of $1$ in $G$, on which $\log=\exp^{-1}$ is a well defined diffeomorphism, such that $g_{1},\ldots,g_{m}\in \mathcal{O}$ generate a dense subgroup if and only if $\log g_{1},\ldots,\log g_{m}$ generate $\mathfrak{g}$.
\end{thm}

Using this theorem leads us to:
\begin{enumerate}[\rm(i)]
\item Prove that we can find suitables powers of the transfer matrices which lies in an arbitrary neighborhood of the identity in $\mathrm{Sp}_{2}(\R)$. These powers will be our ``$g_{1},\ldots,g_{m}$''. To construct these powers we use simultaneous diophantine approximation which can be used only for $E>2$ in our model, as explained in Section $4.1$ of \cite{boumaza}.
\item Compute the logarithms of these powers of transfer matrices. It leads to a first discrete set of $E$'s in $\R$ on which these logarithms are not defined.
\item Out of this discrete set of $E$'s, prove that these logarithms generates the Lie algebra $\mathfrak{sp}_{2}(\R)$ of $\mathrm{Sp}_{2}(\R)$, except for $E$'s in an other discrete subset of $\R$ which corresponds to zeros of some determinants (see Section $4.3$ in \cite{boumaza}). This part of the proof is constructive and for the moment it was not possible to do it for $N$ stricly larger than $2$. 
\end{enumerate}

So finally, in \cite{boumaza}, we were able to prove that there exists a discrete set $\mathcal{S}_{B}\subset \R$ such that for every $E$ in $\mathcal{S}_{B}$, $E>2$, the closed group $G_{\mu_{E}}$ is dense and therefore equal to $\mathrm{Sp}_{2}(\R)$. So we can apply theorem \ref{holderids}, because any compact interval $I\subset (2,+\infty)\setminus \mathcal{S}_{B}$ is also included in an interval $\tilde{I}\subset (2,+\infty)\setminus \mathcal{S}_{B}$ on which $G_{\mu_{E}}$ is $p$-contracting and $L_p$-strongly irreducible for $p\in \{1,2\}$. This finishes the proof of theorem \ref{idshab}.



\end{document}